\documentclass[11pt]{article}
\usepackage[margin=1.25in]{geometry}

\usepackage{graphicx}
\usepackage[sort&compress]{natbib}
\usepackage{amsmath,amssymb} 
\usepackage{booktabs}			
\usepackage[usenames,dvipsnames]{color}    
\usepackage{multirow} 
\usepackage{setspace}  
\usepackage{microtype}
\usepackage{hyperref}
\hypersetup{
    colorlinks,
    linkcolor={Maroon},
    citecolor={MidnightBlue},
    urlcolor={MidnightBlue}
}
\DeclareGraphicsExtensions{.pdf}

\usepackage{bbm}

\newcommand{\HL}{H_{\rm L}}
\newcommand{\HU}{H_{\rm U}}

\newcommand{\Title}{A Statistical Approach to Crime Linkage}

\begin{document}

\newpage
\thispagestyle{empty}
\begin{center}
{\LARGE \Title} \\[20pt]
{\large Michael D. Porter \\[8pt]
\today \\[30pt]
}
\end{center}

\begin{abstract}
The object of this paper is to develop a statistical approach to criminal linkage analysis that discovers and groups crime events that share a common offender and prioritizes suspects for further investigation.
Bayes factors are used to describe the strength of evidence that two crimes are linked. 
Using concepts from agglomerative hierarchical clustering, the Bayes factors for crime pairs are combined to provide similarity measures for comparing two crime series. This facilitates crime series clustering, crime series identification, and suspect prioritization.
The ability of our models to make correct linkages and predictions is demonstrated under a variety of real-world scenarios with a large number of solved and unsolved breaking and entering crimes. 
For example, a na{\"i}ve Bayes model for pairwise case linkage can identify 82\% of actual linkages with a 5\% false positive rate. 
For crime series identification,
77\%-89\% of the additional crimes in a crime series 
can be identified from a ranked list of 50 incidents. 

\vspace{12pt}
\end{abstract}

\vfill
\noindent \rule{0.5\textwidth}{.4pt} \\
\noindent \begin{footnotesize}
Michael D. Porter, Department of Information Systems, Statistics, and Management Science at the University of Alabama, Tuscaloosa, AL (Email: mporter@cba.ua.edu). This work was partially supported by National Institute of Justice Grant 2010-DE-BX-K255. The author expresses gratitude to the Baltimore County Police Department for the crime data.
\end{footnotesize}

\newpage
\onehalfspacing

\section{Introduction}\label{sec:1}
The objective of criminal linkage analysis is to group crime events that share a common offender.
Establishing the set of crimes attributable to a common offender (or set of co-offenders)
is an important first step for several crime modeling techniques. For example, geographic profiling requires the locations from a criminal's crime series in order to predict their anchor point \citep{rossmo}. When interest is focused on predicting the locations of future crimes, the connected series of past events can be used to identify the site selection preferences of 
those committing the crimes, allowing a more precise assessment of future risk \citep{bernasco2005}. 
Establishing a linkage between crimes allows investigators to pool information from various crime scenes, strengthening the case against a serial offender or aiding in offender profiling \citep{Alison-etal-2010}. 
Linkage analysis has also been submitted as fact evidence in legal proceedings \citep{hazelwood03}. 
 
When sufficient physical evidence, such as DNA or fingerprints, is available crimes can be attributed to a specific offender with near certainty. 
In the absence of such evidence it may still be possible to establish, with a high level of confidence, the crimes that were perpetrated by the same individual based on an analysis of the decisions and behaviors of the offender. 
In the course of planning and carrying out a crime, and in response to the situations encountered, the offender will make a series of decisions, perhaps unknowingly, resulting in a set of unique behaviors (i.e., their \emph{modus operandi}).
Linkage analysis uses the measurable results of these decisions (as recorded in a criminal incident database) to determine if crimes are linked.  

The success of linkage analysis rests on three primary assumptions \citep{ grubin2001, Bennell-Jones-2005, Goodwill-Alison-2006, woodhams07case}. The first is that the criminals act consistently. 
While there might be
slight variations in a criminal's behavior as the current goals, situational characteristics, and past learning influence each outcome \citep{Woodhams-etal-2007},
a large body of literature suggests that there is sufficient consistency in some behavioral variables (especially site selection) to allow successful linkage analysis \citep{bernasco08,grubin2001, Bennell-Canter-2002,Bennell-Jones-2005,Goodwill-Alison-2006, Woodhams-Toye-2007, Tonkin-etal-2008,johnson09,Deslauriers-Beauregard-2013,Harbers-etal-2012,Bouhana-etal-2014}.

The second assumption is that offenders exhibit some distinctiveness in their behaviors. If multiple criminals acted identically, there would be no way to distinguish their crimes \citep{Brown-Hagan-2003}. The fact that every person has unique experiences and personality systems suggests that the individual behavior of complex decision makers will exhibit inter-individual variations, even when they are presented with the same situations \citep{shoda1994}. The more variation there is in offender outcomes, the easier a set of crimes can be distinguished from the background \citep{Bennell-Jones-2005}.

The third assumption is that important aspects of the criminals' behaviors can be observed, measured, and accurately recorded \citep{Bennell-etal-2012}.  
This is an important concept because the methods investigators use to observe, measure, record, and code the behaviors can have a profound influence on the potential for linkage. For example, \citet{Tonkin-etal-2008} suggests that intercrime distance provides better linkage performance precisely because it can be more accurately and reliably recorded than the other available crime variables.

The success of establishing linkages is a function of both the similarity and unusualness of the behavioral features recorded from the crime scene \citep{Bennell-etal-2009,Bateman-etal-2007,woodhams07case}.
This suggests that some offenders and crime types may be easier to link than others. Criminals that have unusual, but consistent, behavior (compared to other criminals in the jurisdiction) will be easier to link. Likewise, crime types that exhibit a greater variety of behavioral actions and are less situational or context dependent should be easier to link 
\citep{Bennell-Canter-2002, Bennell-Jones-2005, Woodhams-etal-2007}.

The term crime linkage can refer to several connected, but slightly different tasks.  
We refer to \emph{pairwise case linkage} as the processes of determining if a given pair of crimes share the same offender. This is a binary classification problem where each crime pair is considered independently. 
There have been numerous studies detailing the performance of case linkage methods across a variety of crime types \citep[e.g.,][]{Woodhams-Labuschagne-2012,Markson-etal-2010,Tonkin-etal-2011,Tonkin-etal-2012,Bennell-Jones-2005,Goodwill-Alison-2006,Woodhams-Toye-2007,Bennell-etal-2009,Tonkin-etal-2008,Brown-Hagan-2003,Lin-Brown-2006,Cocx-Kosters-2006}.

Expanding consideration beyond crime pairs, \emph{crime series identification} 
is the process of identifying all of the crimes (i.e., the crime series)
that were committed by the same offender. 
One approach, termed reactive linkage in \citet{woodhams07case}, starts with an index crime or set of crimes and attempts to discover the additional crimes that share a common offender with the index crime(s).
This could be used to generate a list of other crimes that a suspect may have committed for investigative or interrogative purposes.
A variety of methods have been proposed for reactive linkage \citep[e.g.,][]{Adderly-Musgrove-2003,Adderly-2004,Wang-etal-2013,Reich-Porter-CrimeClust}.
The other approach, termed proactive linkage in \citet{woodhams07case} and what we refer to as \emph{crime series clustering}, treats crime linkage as a clustering problem and attempts to cluster all of the crimes in a criminal database such that each identified cluster corresponds to a crime series. 
A variety of clustering methods have been used to group similar crimes for further investigation or offender profiling \citep[e.g.,][]{Adderly-Musgrove-2001,Ma-etal-2010,Reich-Porter-CrimeClust}.

Because many crimes are due to repeat offenders \citep{Langan-Levin-2002}, linkage analysis could be important for \emph{suspect prioritization}.
This task attempts to link crimes to offenders by comparing an unsolved crime or crime series
to the crimes known to be perpetrated by a set of past offenders.
Several supervised learning (i.e., classification) methods have been employed to generate a ranked list of suspects for further investigation \citep[e.g.,][]{Canter-Hammond-2007, Ewart-etal-2005,Snook-etal-2006,Santtila-etal-2004,Santilla-etal-2005,Santtila-etal-2008,Salo-etal-2012,Yokata-Watanabe-2002,Winter-etal-2013}. 

The purpose of this paper is to introduce criminal linkage analysis from a statistical perspective. 
Based on Bayesian decision theory, we consider the use of \emph{Bayes Factors} to directly assess the similarity and distinctiveness of a pair of crimes. 
The resulting model is in the form of a posterior odds ratio (related to the likelihood ratio), with the numerator estimating the joint probability of observing two crime event variables given a common offender (similarity) and the denominator estimating the probability given different offenders (distinctiveness). The Bayes factors provides a convenient multiplicative form for quantifying the evidence, individually and jointly, that two crimes are linked.
We show how some models previously used for case linkage (e.g., logistic regression, na{\"i}ve Bayes) estimate the Bayes factor by assuming it is of a special form.

Furthermore, we present a new methodology for crime series identification and clustering by combining the Bayes Factors from pairwise comparisons with methods from hierarchical clustering to link sets of crimes.
This allows the many existing models for case linkage to be used for crime series identification, clustering, and suspect prioritization. 
To illustrate the practical use and performance of our methodology, these various types of linkage are applied to a large criminal incident dataset that includes both solved and unsolved crimes. 

\section{Evidence Variables} \label{sec:transformations}
One of the initial steps in case linkage involves the collection and preprocessing of the data corresponding to each crime event \citep{woodhams07case}. 
This may involve extracting and coding offender behaviors or physical descriptions from an incident report, geocoding the address of the crime location, or processing forensic evidence. 
Let $V_i$ be the resulting vector of crime variables for crime $i$. 
This crime data will contain information regarding the offender (if available), crime, and crime scene (e.g., spatial location, timing, crime type, offender behavior, victim characteristics). 
As such, the crime data will likely consist of a combination of categorical, discrete, continuous, and missing valued variables.

After the relevant crime data are extracted and coded, case linkage compares the variables from two crimes to assess their level of similarity and distinctiveness.
To facilitate this comparison, the crime variables for a pair of crimes can be transformed and converted into \emph{evidence variables} 
that are more suitable for linkage analysis. 
Let $X_{i,j} =[X_1(i,j),X_2(i,j),\ldots,X_p(i,j)]$ denote the $p$ evidence variables extracted from the data of crimes $i$ and $j$. 
The $m^{th}$ evidence variable $X_m(i,j)=T_m(V_i,V_j)$ is obtained by applying a transformation function $T_m$ to the crime data. 
The transformation functions can take a variety of  forms depending on the nature of the data and type of variable, but we assume that they are invariant to the order of the crime variables (i.e. $T(V_i,V_j)=T(V_j,V_i)$). 

The transformation functions can produce measures of similarity or dissimilarity.
For example, spatial distance is an appropriate transformation for comparing the dissimilarity of two crime locations \citep{Bennell-Canter-2002}.
Temporal dissimilarity can be represented by the number of days between crimes  \citep{Goodwill-Alison-2006}. 
A binary measure of similarity for categorical crime variables is the indicator function \citep{Brown-Hagan-2003}:
$T(V_i,V_j)=1$ for matching categories ($V_i=V_j$) and $T(V_i,V_j)=0$ for non-matching categories ($V_i \neq V_j$).

There are many additional transformation functions that are appropriate for linkage data (e.g., groups of behavioral crime variables can be compared with Jaccard's coefficient \citep{Melnyk-etal-2011} or a matching index \citep{Ellingwood-etal-2012}).
The selection of appropriate transformation functions is an important step in the linkage process as different transformations will provide different linkage performance \citep{Melnyk-etal-2011,Bennell-etal-2010,Ellingwood-etal-2012,Salo-etal-2012} and can impact the class of models available for linkage.
The next section shows how the resulting evidence variables are used to estimate the Bayes factor.

\section{Statistical (Pairwise) Case Linkage}\label{sec:methods}
This section formally introduces a statistical approach to pairwise case linkage and proposes the use of the Bayes factor for making decisions regarding linkage. The (log) Bayes factor can be thought of as the weight of evidence \citep{Good-1985,Good-1991}; a measure of how strong the evidence favors the linkage hypothesis.
When the number of useful evidence variables is small, it might be possible to calculate the Bayes factor directly. 
However, when this is not the case we show how the Bayes factor can be estimated (and important variables selected) using several common methods for binary classification (e.g., na{\"i}ve Bayes, logistic regression, Bayesian networks, random forests, and boosting).

\subsection{Bayesian Case Linkage}
We define criminal case linkage as the task of determining whether two crimes share a common offender.  
As such, case linkage only considers crime pairs for linkage.
Extensions for linking a set of crimes (i.e., crime series linkage) is given in Section~\ref{sec:seriesLinkage}.

Our approach is to cast the case linkage problem in the form of a Bayesian hypothesis test comparing the estimated probability of the linkage hypothesis $\HL$ to the unlinked hypothesis $\HU$ for each pair of crimes.
For case linkage, the posterior probability that crimes $i$ and $j$ are linked is given by
\begin{align}
\Pr(\HL \mid X_{i,j}) = \frac{\Pr(X_{i,j} \mid \HL)\Pr(\HL)}{\Pr(X_{i,j})} 	
\end{align} 
where $X_{i,j}$ represent the extracted evidence variables (see Section \ref{sec:transformations}) from crimes $i$ and $j$, $\Pr(\HL)$ is the prior probability that the two crimes are linked, and $\Pr(X_{i,j} \mid \HL)$ is the likelihood of observing the data under the linkage hypothesis.
The two competing hypotheses can be compared via the posterior odds
\begin{equation} 
\underbrace{ \frac{\Pr(\HL \mid X_{i,j})}{\Pr(\HU \mid X_{i,j})}}_{\text{ \rm Posterior Odds}} = 
\underbrace{\frac{\Pr(X_{i,j}\mid \HL)}{\Pr(X_{i,j}\mid \HU)}}_{\text{\rm Bayes Factor}} \,\times \underbrace{\frac{\Pr(\HL)}{\Pr(\HU)}}_{\text{ \rm Prior Odds}  }
\label{eq:posterior.odds},
\end{equation}
which is a measure of the extent to which the data and prior specifications support the linkage hypothesis. 

The first term on the right-hand side of \eqref{eq:posterior.odds} is the \emph{Bayes factor} \citep{Kass-Raftery-1995,Raftery-1995}.
The Bayes factor has an attractive form: the numerator can be viewed as a measure of the similarity between crimes and the denominator a measure for their distinctiveness. 
The numerator, $\Pr( X_{i,j} \mid \HL)$, is the likelihood of observing the crime data when the crimes are indeed linked. Thus, under the assumption of offender consistency \citep{canterD95}, this will be large when both crimes have similar characteristics. 
On the other hand, the denominator $\Pr( X_{i,j} \mid \HU)$ is the likelihood of observing the data when the crimes were committed by different offenders. 
This will take large values whenever the characteristics of the two crimes are common amongst all offenders. For example, if 50\% of all burglaries involved the use of force to gain entry, then it would not be too surprising (i.e., 25\%) to find a pair of unlinked burglary crimes that also used force as a method of entry.
Thus, the Bayes factor offers a formal and explicit way to measure the similarity between crimes while accounting for their distinctiveness, a crucial requirement of linkage analysis.

\subsection{Properties of the Bayes Factor}
By rearranging \eqref{eq:posterior.odds}, the Bayes factor can be viewed as an odds ratio: the ratio of posterior odds to prior odds.
As such, the Bayes factor describes how much additional strength is given to the linkage hypothesis after observing the crime data.
Based on this property, \citet{Good-1985,Good-1991} defines the log of the Bayes factor as a ``weight of evidence" (in favor of the linkage hypothesis). 

The Bayes factor for a pair of crimes $i$ and $j$,
\begin{equation}
BF(i,j) = \frac{\Pr(X_{i,j}\mid \HL)}{\Pr(X_{i,j}\mid \HU)},
\label{eq:bayes.factor}
\end{equation}
will take a value between 0 (when it is assessed impossible to observe the data $X_{i,j}$ if the crimes are linked) and $\infty$ (when it is assessed impossible to observe the data if the crimes are unlinked). 
The Bayes factor can be interpreted as the magnitude with which the crime data favors the linkage hypothesis, e.g., a Bayes factor of 5 implies that the linkage hypothesis is supported 5 times as much by the crime data than the unlinked hypothesis.

Another benefit of using the Bayes factor instead of the posterior odds or posterior probabilities is that it is independent of the prior probabilities. The Bayes factor can be directly compared across studies or samples, while comparisons using the posterior probabilities will need to adjust for different prior probabilities. 
Also, the prior probability of a linkage will be difficult to estimate in practice when unsolved crimes are involved in the analysis. 

Furthermore, decision making concerning a linkage can be based directly on the Bayes factor.
According to statistical decision theory \citep{Berger-1985}, it is optimal to declare a linkage when the Bayes factor exceeds a threshold based on the prior odds and cost of wrong decision, i.e.,
\begin{align}
BF \geq \frac{\Pr(\HU)}{\Pr(\HL)} \times \frac{\text{Cost of false linkage claim}}{\text{Cost of missed linkage}}
\label{eq:BFthres}
\end{align}
This permits decisions based on the costs of making an erroneous linkage decision (e.g., it may be more costly to miss a linkage than issue a false alarm). 
When the ratio of costs or prior odds are not known, or hard decisions are not necessary, receiver operating characteristic (ROC) curves \citep{Bennell-Jones-2005,Bennell-etal-2009} can be constructed from the Bayes factors. These give a visual representation of the trade-off between the false alarm and true detection rate for every possible threshold.

\subsection{Estimating the Bayes Factor}
\label{sec:BFmodels}
When the number of evidence variables is small, direct estimation of the Bayes factor may be possible. 
However, as the number of evidence variables grows, simplifying models may facilitate better estimations.
The selection of a useful modeling strategy can be assisted by the recognition that pairwise case linkage is essentially a binary classification problem \citep{Bennell-Canter-2002}.
Taking this view, we examine how various classification methods estimate the Bayes factor for case linkage.

A na{\"i}ve Bayes model makes the simplifying assumption that the evidence variables are independent. 
Under this assumption, the joint Bayes factor becomes the product of the individual Bayes factors (i.e., 
$BF = BF_1 \times BF_2 \times \cdots \times BF_p$ or equivalently $\log BF=\log BF_1 + \log BF_2 + \cdots + \log BF_p$).
Because this assumption requires that only the univariate Bayes factors be calculated, the computational complexity is reduced significantly.
While the assumption of independence is likely to be violated, the resulting model can be quite competitive, especially when there are many evidence variables \citep{Hand-Yu-2001}.		
A na{\"i}ve Bayes classifier was used for suspect identification in \citet{Salo-etal-2012} and \citet{Winter-etal-2013} using binary behavioral variables.

Logistic regression models the log posterior odds of a linkage as a linear function of the evidence variables,
\[\text{log(Posterior Odds)}= \phi + \beta_0+\beta_1 X_1 + \cdots + \beta_p X_p\]
where $X_m \,(m =1,2,\ldots, p)$ is the $m^{\rm th}$ evidence variable for a pair of crimes and $\phi$ is an estimate of the log prior odds.
Consequently, 
$\log BF = \beta_0+\beta_1 X_1 + \cdots + \beta_p X_p$
is the logistic regression estimate of the log Bayes factor.
Logistic regression was introduced for case linkage by \citet{Bennell-Canter-2002} and subsequently employed in the majority of case linkage studies. 

While na{\"i}ve Bayes and logistic regression are popular models for classification, they make strong assumptions about how the evidence variables relate to the Bayes factor.
More complex models can offer less restrictive assumptions.
For example, Bayesian networks eliminate the independence assumption of na{\"i}ve Bayes by attempting to identify and model the multivariate dependency structure between the evidence variables. 
This requires more training data and potentially simpler functions for the conditional Bayes factors. 
Bayesian networks have been used to predict offender profiles from unsolved crimes \citep{Baumgartner-etal-2008,Ferrari-etal-2008}, but not to our knowledge for case linkage.

Logistic regression can be extended by regularization methods (i.e., penalized regression) that estimate the coefficients subject to a constraint on their magnitude \citep{Friedman-etal-2010,Hastie-etal-2009}.
This can provide better variable selection (compared step-wise approaches) and better prediction.
Another extension is to relax the linear form that standard logistic regression imposes on the Bayes factors by using generalized additive models (GAM) \citep{Hastie-Tibshirani-1990} or boosting \citep{Friedman-etal-2000}. 
To facilitate complex interactions between the evidence variables, classification trees \citep{Tonkin-etal-2012}, random forests, gradient boosting, and other ensemble methods can be employed \citep{Berk-2006,Berk-2013}.

To summarize, there are four primary steps for Bayesian case linkage.
First, transformation functions ($\{T_m\}_{m=1}^p$) are established to convert the raw crime data for a pair of crimes into evidence variables.
Secondly, a predictive model is selected (e.g., na{\"i}ve Bayes, logistic regression, boosting). 
This may require specifying distributions (and hyperparameters) for the evidence variables. 
Next, training data (possibly augmented with expert opinion) is used to update knowledge about the unknown model parameters. 
Lastly, the resulting (fitted) Bayes factor model is output to allow comparison of new crime pairs.
The first three steps could be based on an initial exploratory analysis of the training data. 
The last step outputs a model $f$ such that $\widehat{BF}=f(X_1,X_2\ldots,X_p)$ predicts the Bayes factor for crime pairs that were not included in the training data.

\section{Crime Series Linkage} 
\label{sec:seriesLinkage}
Pairwise case linkage attempts to establish if a \emph{pair} of crimes share a common offender.
In practice, there is more interest in crime series linkage, which attempts to identify the \emph{set} of crimes committed by a common offender.
For example, a crime analyst may need to identify the additional crimes that are part of a crime series (\emph{crime series identification}), create a ranked list of suspects for a particular set of crimes (\emph{suspect prioritization}), or discover all of the crime series in a criminal incident database (\emph{crime series clustering}).
Our approach to crime series linkage involves creating measures of similarity between sets of crimes that are functions of the pairwise Bayes factors obtained from case linkage. 
The sets of crimes are then linked using methods from hierarchical cluster analysis to identity the individual crime series.

\subsection{Hierarchical Clustering}
Hierarchical clustering is an algorithmic approach to cluster analysis that sequentially forms a hierarchy of cluster solutions \citep{Borgatti-1994,Mullner-2011}. 
The agglomerative approach starts with every observation (e.g., crime incident) in its own cluster. Then it sequentially merges the two closest clusters to form a new larger cluster. This process is repeated until all observations are in the same cluster or a stopping criterion is met. 
This algorithm requires two similarity measures to be specified: the pairwise similarity between two observations and the similarity between two groups of observations.
Equivalently, dissimilarity or distance measures can be used instead of similarity measures with the obvious adjustments to the algorithm.
There are three primary approaches to measuring the similarity between groups of observations. 
\emph{Single linkage}, or nearest neighbor, uses the most similar pair between the two groups as the group similarity measure. In contrast, \emph{complete linkage} uses the least similar pair between two groups as the measure of group similarity. \emph{Average linkage} uses the average similarity between all pairs in the two groups. 

\subsection{Agglomerative Crime Series Clustering}
Crime series clustering is fundamentally a clustering problem where we seek to group the crimes into clusters that correspond to a serial offender.
We use the log Bayes factor as the pairwise similarity measure between all unsolved crimes. That is, the similarity between crimes $i$ and $j$ is $S(i,j) = \log BF(i,j)$, where $BF(i,j)$ is the estimated Bayes factor for linkage. 
We choose to use the Bayes factor rather than the posterior probability (or posterior odds) to remove dependence on the prior probability $\Pr(\HL)$. 
This will not make a difference if single or complete linkage is used as these approaches only rely on the ordering of the similarity scores. However, different clustering results could be obtained if posterior probabilities are used with the average linkage clustering.

Once the pairwise similarity between all unsolved crimes has been calculated, a measure for cluster similarity must be constructed. The clusters in hierarchical clustering will consist of either singleton crime incidents or groups of crimes. In this framework, a group of crimes is assumed to be a crime series sharing the same offender(s).   
Thus, the cluster similarity measure should reflect how likely it is that the crimes in two clusters have the same offender. 
The initial iterations in the hierarchical clustering algorithm will consist of merging single crime incidents with single crime incidents (\emph{incident-incident similarity}). In this special case, all three linkage methods (single, complete, and average) give the same result. 
However, the three methods will differ on how they specify the \emph{incident-series similarity} and \emph{series-series similarity} measures.

By using the most similar pair between clusters to define the similarity, the single linkage approach has the tendency to link together long chains of crimes. This has the potential to create clusters in which some crimes are very dissimilar. While this is usually viewed as a disadvantage of single linkage, for crime modeling it may be advantageous. 
For example, this could detect series from offenders who are slowly changing their preferences or series from multiple co-offenders who sometimes act together and sometimes separately. 
Complete linkage can avoid the long chains by using the least similar pair to measure the similarity between two clusters. This can better discriminate the series arising from a particular group of co-offenders. While the average linkage method can be viewed as a compromise between single and complete linkage, it does 
provide different results if the posterior odds or posterior probability is used in place of the log Bayes factor for the pairwise similarity measure.

\begin{figure}
\centering
\includegraphics[height=.85\textwidth,angle=-90]{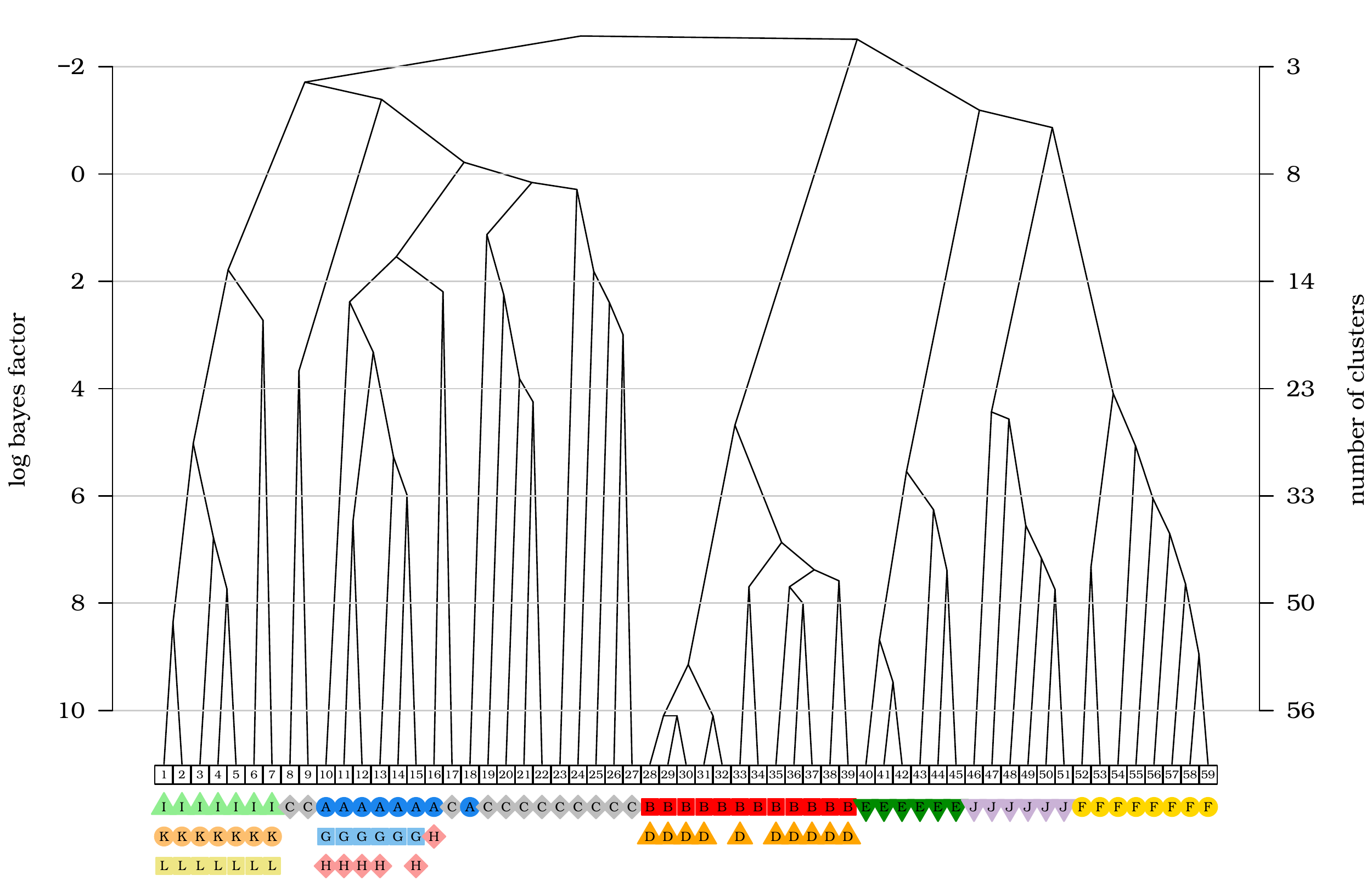}
\caption{Dendrogram (using average linkage) for the 59 crimes from 12 offenders (labeled A-L) who committed at least 6 crimes. Below the crime number is the label (A-L) of the offender or set of offenders arrested for the crime.}
\label{fig:dendroSeries}
\end{figure}

Hierarchical clustering produces a collection of clustering solutions that starts with each crime in its own individual cluster and ends with all crimes in one single cluster. 
Dendrograms (Figure~\ref{fig:dendroSeries}) are graphical representations of the clustering hierarchy showing the entire solution path and cluster similarity scores. 
Cutting the dendrogram at a given similarity threshold partitions the crimes into disjoint clusters that estimate the crime series present in the data. 
Consider the  dendrogram in Figure~\ref{fig:dendroSeries} representing the results from average linkage agglomerative clustering for the 59 crimes from 12 offenders. 
Each crime is represented by one leaf node at the bottom of the dendrogram. Below the crime number is the label (A-L) of the offender or set of offenders arrested for the crime (several crimes have multiple offenders).
Notice that crimes 29 and 30 were the most similar (i.e., had the largest Bayes factor) and were merged first. Because both of these crimes were also very similar to crime 28, the three crimes were grouped together with a score of greater than 10. All of Offender B's crimes (with or without Offender D's involvement) were clustered with a score of around 5 suggesting that this offender has offended in a relatively consistent and/or distinctive manner. 
Contrast this with the crimes from Offender C. While some of their crimes are linked with a score around 4, the full crime series is not connected until the score becomes negative and at would include the crimes from Offenders A, H, and G. 
Consider the crime series involving Offender A. When co-offending with Offenders H and G, the crimes can be linked with a relatively large score (around 2.5), but when Offender A acts alone (crime 18) or only with Offender H (crime 16) the crimes are clustered into a group with Offender C. 

Because the cluster similarity score is based on the log Bayes factor, 
the score at which two clusters are merged provides insight into the strength of similarity, allowing a crime analyst to act only when appropriate. Sensible thresholds for determining a linkage can be determined from a variety of methods. 
For example, setting a threshold of 0 would only merge groups when the data favors a linkage between at least one pair of crimes. 
Furthermore, an appropriate threshold could be derived from equation~\eqref{eq:BFthres} or performance on training data. 

The dendrogram also facilitates investigations about particular crimes. 
Starting with a single crime, a path can be traced up the dendrogram to find all of the other crimes that may be part of its series (see Section \ref{sec:csi} for a more detailed treatment of this concept). 
This could be used to generate a list of other crimes that a suspect may have committed for investigative or interrogative purposes.

Figure~\ref{fig:AHC} shows the algorithm for agglomerative crime series clustering with Bayes factors.
Before analysis, some 
crimes may already be grouped into series through other forensic evidence (e.g., DNA), suspect descriptions, or by crime analysts. This information can be incorporated into the hierarchical clustering approach 
by initiating the algorithm with the predetermined clusters and proceeding as usual. This is commonly termed semi-supervised clustering.

\begin{figure}[h]
\hrulefill \\
\textbf{Agglomerative Hierarchical Clustering Algorithm for Crime Series Clustering} \\[10pt]
\textbf{Initialize:}\\
1.  Calculate the similarity between all crime incidents using the log Bayes factor\\
2.  Assign every unsolved crime to its own cluster (or into predefined clusters if available)\\
\textbf{Iterate:}\\
3.  Calculate the similarity between all clusters (using single, complete, or average linkage)\\
4.  Merge the two most similar clusters into one new cluster\\
5.  Repeat 3-4 until there is a single cluster or stopping criterion is met \\
\hrule 
\caption{Hierarchical clustering algorithm for crime series clustering}
\label{fig:AHC}
\end{figure}

\subsection{Crime Series Identification} \label{sec:csi}
Instead of simultaneously clustering all crimes in a criminal database, crime series identification is focused on identifying the additional crimes that are part of a specific existing crime series.
Crime series identification can be useful in interrogations \citep{Adderly-Musgrove-2003,Adderly-2004} by providing investigators a list of additional crimes that a suspect may be responsible for. 

A crime analyst could also use crime series identification when investigating a crime series or suspected crime series to generate a list of additional crimes to investigate.

Assume that $C = \{V_1,\ldots,V_n\} \subset \mathcal{C}$ is a crime series (or single crime) of interest and the goal is to find all of the additional crimes in a criminal incident database $\mathcal{C}$ that are attributable to the same offender(s) responsible for the crimes in $C$. 

Our approach is to compare every unsolved crime in $\mathcal{C} \setminus C$ to $C$ using single, complete, or average linkage (with the log Bayes factor as the similarity score). 
A ranked list can be provided to investigators of the crimes most similar to those in $C$ along with the Bayes factor based scores. 

If it can be assumed that there are many additional crimes in the series, a clustering approach may be advantageous. 
For example, the results from crime series clustering (i.e., clustering all crime events) can identify similar crimes by tracing a path up the dendrogram.
A similar approach starts with $C$ in a cluster by itself and adds crimes sequentially according the linkage method (e.g., single, complete, average). 
A potential advantage of the clustering based approaches is that each additional crime changes the cluster and can affect the subsequent crimes added. This can help investigators find additional crimes attributable to an offender even if their tactics shifted over the course of offending.

\subsection{Suspect Prioritization} 
The goal of suspect prioritization is to assist the police in identifying the offender of 
a set of unsolved crimes by providing a ranked list of suspects.
\citet{Snook-etal-2006} combines aspects of journey-to-crime and criminal career analysis by prioritizing based on the distance between a crime and the current residence of certain past offenders.
\citet{Canter-Hammond-2007} extends this concept by using geographic profiling methods to incorporate multiple crimes.
When the past offenders' current home location is unavailable or deemed uncertain, the characteristics of their past crimes can be used to assess how likely they are responsible for the current crime(s) under consideration \citep{Yokata-Watanabe-2002}. This is the approach we consider.

This view of suspect prioritization is very similar to crime series identification except instead of comparing an unsolved crime series $C \subset \mathcal{C}$ to every unsolved crime in $\mathcal{C} \setminus C$, it is compared
to the crime series from past offenders (i.e., the \emph{solved} crimes in $\mathcal{C}$).
If $\{C_m \subset \mathcal{C}:m=1,2,\ldots,M\}$ is the set of crime series attributable to the $M$ past offenders, the task of suspect linkage is to compare $C$ to every $C_m$ and rank suspect $m$ by how similar their past crime series is to $C$.
In practice, because we are using the crime series from past offenders (which may be from the distant past), we do not include temporal distance in calculating the Bayes factor.

\section{Analysis of Data} \label{sec:analysis}
\subsection{Data Description}

The Baltimore County Police Department provided data on $n=$ 10,675 breaking and entering crimes reported in 2001-2006. Each crime record includes a geocoded location, the time interval in which the crime could have occurred,
the type of property (34 levels), the point of entry (8 levels), the method of entry (16 levels), and an anonymized offender identifier for the solved crimes. 

From these crime variables, we considered seven evidence variables for comparing a pair of crimes. 
Four are continuous valued and three are binary.
Spatial distance (\texttt{spatial})  measures the Euclidean distance between crimes,
temporal proximity (\texttt{temporal}) measures the elapsed time between crimes, time-of-day distance (\texttt{tod}) measures the difference in the time of day between crimes (regardless of how many days apart the two crime occurred),
and day-of-week distance (\texttt{dow}) measures the difference in day of week between crimes (regardless of how many weeks apart the two crime occurred).
The indicator function was used to create three binary evidence variables:
property type (\texttt{prop}), point of entry (\texttt{poe}), and method of entry (\texttt{moe}). Each of these variables have the value 1 when the categories match and 0 if they do not match.
 
If a crime occurs when the victim or a witness is not present, the exact timing of the crime may not be available.
For such crimes, victims indicate the likely interval of time during which the crime could have occurred (i.e., the event times are interval censored). 
Because breaking and entering crimes often occur when the victim is away from their property, uncertainty is prevalent in the event timing. 
Only about 26\% of crimes have an exact time recorded and 55\% have an interval shorter than 6 hours. 
Because there is so much uncertainty, using only one point (e.g., earliest time, midpoint, latest time) to represent the event time could result in too much bias \citep{Ratcliffe-2002}.
To help account for the uncertainty, our transformation function for the temporal variables calculates the \emph{expected} differences.
For each crime pair, the expected differences are estimated by simulating 1000 values for uncertain event times from a uniform distribution over its interval and taking the average absolute differences.

We only consider crimes of a single type. 
While recent work has shown that linkage can successfully incorporate multiple crime types \citep{Tonkin-etal-2011}, it has only considered space-time variables. In this situation, a categorical crime type variable could be added in a straightforward manner.
However, including other crime variables (that are not similar across crime types) would make the model very complex and specialized approaches would have to be considered.

\subsection{Training Data}
The crime data are partitioned into a training period (2001-2005) to allow exploratory analysis and model building and a testing period (2006) to provide an evaluation of our methods.
The training data is composed of all solved crimes occurring from 2001-2005. 
A crime is considered solved if a suspect was arrested.
There were a total of 4681 solved breaking and entering crimes during this time period attributed to a total of 4147 detected criminals. While 3375 (81\%) of detected criminals were attributed a single crime, 772 (19\%) had a series of at least 2 crimes and 149 (4\%) were charged with a series of at least 5 crimes. 
There is also a large amount of co-offending with 1142 (24\%) of the crimes involving multiple offenders. 

To construct models for pairwise linkage, we need to generate two sets of crime pairs (linked and unlinked) from the training data.
However, these data present two challenges in creating an appropriate set of crime pairs: unequal series length and co-offending.
Because a crime series of length $n$ generates $N=n(n-1)/2$ linked pairs, a long series will contribute substantially more linked pairs than a short series. 
A common approach to alleviate the potential bias is to form linked pairs from only a small number (e.g., 2) of crimes from each series \citep{Bennell-Canter-2002,Markson-etal-2010}. 
While this approach does treat each series equivalently, it reduces the amount of data used to build our models \citep{Deslauriers-Beauregard-2013}. 
Instead, we use all $N$ linked pairs from a crime series, but weight each pair by $1/N$. 
Due to the presence of co-offending, some crimes will be present in multiple series. 
If not accounted for, this could generate the same linked pair multiple times (but with potentially different weights).
To prevent this, we only use the unique linked pairs and assign the smallest weight from all replications. 
Formally, the weight for a linked crime pair $V_i,V_j$ is $\min \{1/N_m:V_i,V_j \in C_m \}$, where $C_m$ is the crime series for offender $m$ and $N_m=|C_m|$ is the number of crimes in their series (assuming $V_i$ and $V_j$ are in at least one crime series).
While this approach necessitates models that can handle weighted observations\footnote{An alternative is to sample according to the weights.}, it ensures that all of the data are considered while also maintaining equal treatment for every series.

We also need to consider co-offending and series length in how we select the crime pairs to generate the unlinked pairs. 
To account for co-offending, we first make an offender graph that connects two offenders if they have co-offended during the training period.
Then we extract the \emph{crime groups} as the group of offenders who can be connected (i.e., the connected components of the offender graph).
Because this associates each crime with only one crime group, we can consider crimes from different crime groups to be unlinked. 
Although not all crimes within a crime group will be linked (e.g., some crime pairs within a group may have different offenders), this approach increases the chances that two crimes taken from different crime groups are indeed unlinked.
To generate the unlinked pairs, we select 20 crimes (with replacement) from each crime group and pair them with 20 crimes from different groups. Duplicates are removed and each unlinked pair is assigned a weight of one. 

Additionally, the time between crime pairs in the training set is restricted to be no more than $365$ days. 
This step is taken for several reasons. 
The test set spans only one year,  
so crime pairs separated by more than this will not be evaluated. 
Because some Bayes factor models (e.g., logistic regression) can be adversely impacted by outliers and extreme values, limiting the time period has the potential to improve model performance.
Also, long stretches between linked crimes may indicate that an offender was incarcerated or not residing in the same area. If they do re-offend, it is more likely that they have different preferences (e.g., they live in a different location or learned different criminal skills). 
We could also consider a similar spatial constraint \citep{Burrell-etal-2012}, but didn't in this analysis because Baltimore County was deemed sufficient for an offender to traverse by vehicle if they desired.

This approach leads to $11{,}077$ pairs of linked crimes (with weights) and $42{,}369$ pairs of unlinked crimes. 
Each pair is then transformed into evidence data and a linked or unlinked label is attached.

\subsection{Model Building} 
Three models were considered for estimating the Bayes factor: logistic regression, na{\"i}ve Bayes, and boosted trees. 
The estimated parameters from logistic regression are given in Table~\ref{t:logreg}.
Spatial and temporal distances (including time-of-day and day-of-week) have a significant negative coefficient implying that ``closer" crime pairs are more likely to be linked.
Matching property type, method of entry, and point of entry increases the evidence of a linkage (with a matching property type variable providing the strongest indication). 

\begin{table}[!ht]
\caption{Logistic regression parameter estimates for case linkage.}
\label{t:logreg}
\begin{small}
\begin{center}
\begin{tabular}{rrrrr}
 \toprule
 & Estimate & SE & $z$-val & $p$-val \\ 
 \midrule
(Intercept) & -0.14 & 0.16 & -0.93 & 0.35 \\ 
  spatial & -0.31 & 0.01 & -22.61 & 0.00 \\ 
  temporal & -0.01 & 0.00 & -17.82 & 0.00 \\ 
  tod & -0.10 & 0.02 & -5.91 & 0.00 \\ 
  dow & -0.24 & 0.05 & -4.79 & 0.00 \\ 
  prop.match & 1.25 & 0.10 & 12.42 & 0.00 \\ 
  poe.match & 0.41 & 0.10 & 4.13 & 0.00 \\ 
  moe.match & 0.28 & 0.12 & 2.43 & 0.02 \\ 
  \bottomrule
\end{tabular}
\end{center}
\end{small}
\end{table}

\begin{figure}[!ht]
\centering
\includegraphics[width=\textwidth]{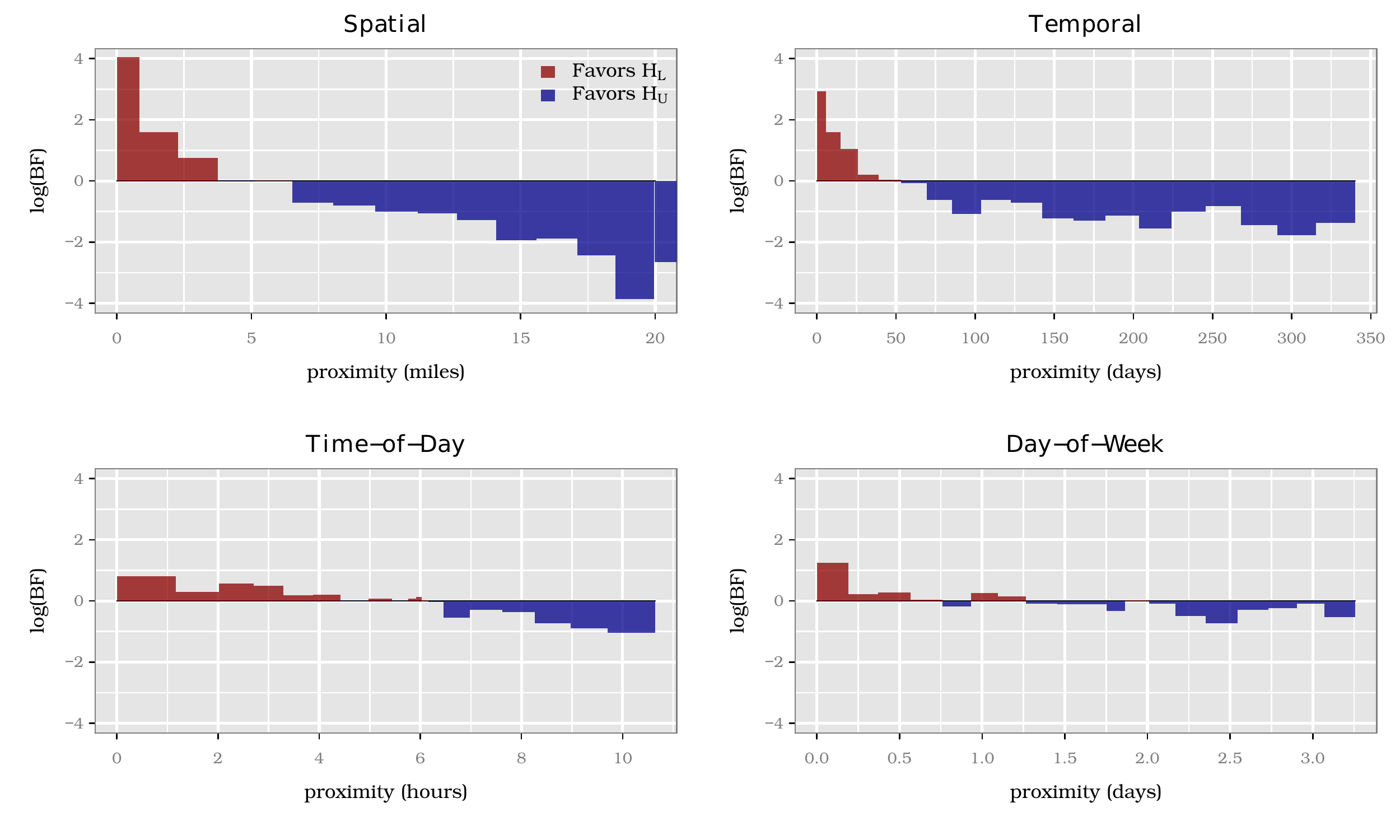} \\
\includegraphics[width=\textwidth]{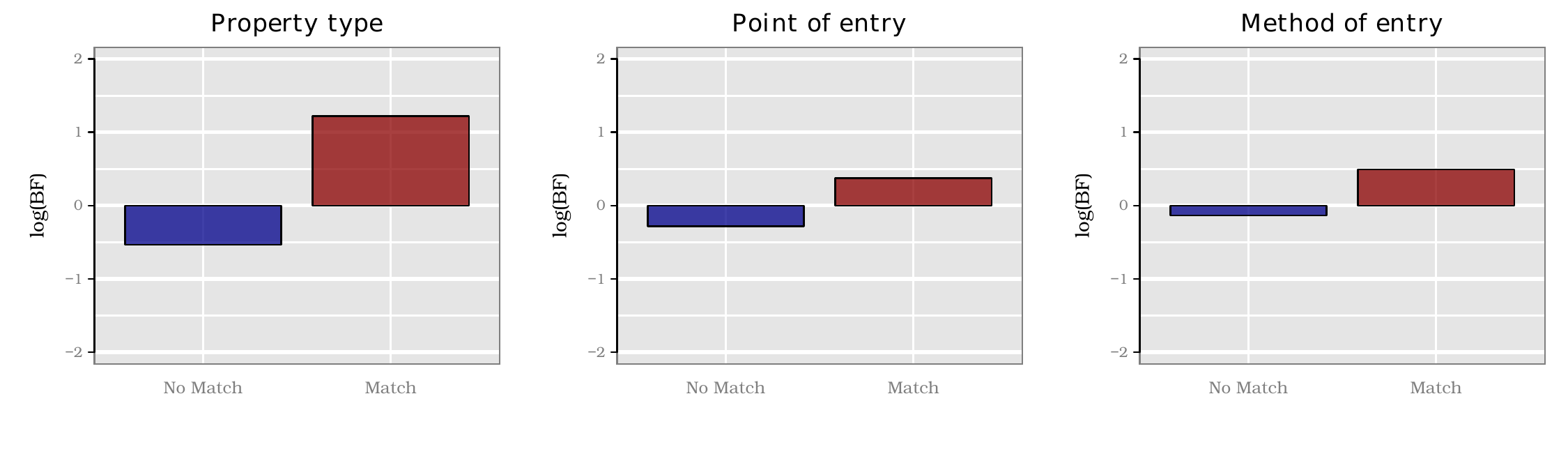}
\caption{Component plots of the estimated log Bayes factor.}
\label{fig:BF}
\end{figure}

For the na{\"i}ve Bayes model, the continuous variables were discretized into 20 bins according to equal frequency binning. 
The estimated (component) Bayes Factor for variable $j$ taking a value in bin $k$ is
\begin{equation}
BF_{j}(k) = \frac{l_{j}(k)/L}{u_{j}(k)/U}
\end{equation}
where \{$l_{j}(k)$, $u_{j}(k)$\} are the sum of the weights for the \{linked, unlinked\} observations with values in bin $k$ and \{$L$, $U$\} are the total weights for all \{linked, unlinked\} observations.
Figure~\ref{fig:BF} shows the component plots for all the predictors incorporated in the model.
The spatial and temporal proximity variables have the largest range in log Bayes factors indicating that they are most influential. 

A gradient boosted tree model \citep{Ridgeway-gbm} was also fit using trees with 4 levels of interaction. 
The number of iterations (i.e., stopping) was determined from out-of-bag performance. 
Table~\ref{t:varImportance} shows the normalized variable influence score \citep{Friedman-2001,Ridgeway-gbm}. 
For this model, spatial and temporal distances were the most important predictors.
Because of the black-box nature of boosted tree (with interactions), it is not clear how each predictor affects the linkage decision.
\begin{table}[ht]
\centering
\caption{Variable influence scores from boosted trees} 
\label{t:varImportance}
\begin{small}
\begin{center}
\begin{tabular}{lr}
  \toprule
var & rel.inf \\ 
	\midrule
spatial & 71.50 \\ 
  temporal & 23.06 \\ 
  tod & 2.38 \\ 
  prop & 1.81 \\ 
  dow & 0.83 \\ 
  poe & 0.38 \\ 
  moe & 0.04 \\ 
  \bottomrule
\end{tabular}
\end{center}
\end{small}
\end{table}

\subsection{Testing Data}
A test set was constructed from the 2006 data.
There were $946$ solved crimes and $5048$ unsolved crimes. The solved crimes were committed by $1047$ offenders with an average series length of $1.3$. 
Table~\ref{t:seriesLength} shows frequency of crime series lengths for the known crime series from the testing period.
There is also a substantial amount of co-offending during the testing period with 31.5\% of the solved crimes having multiple offenders.
\begin{table}[ht]
\begin{center}
\caption{Crime series length for crimes in testing period}
\label{t:seriesLength}
\begin{small}
\begin{tabular}{r|lllll}
\toprule
\# crimes & 1 & 2 & 3 & 4 & 5+ \\ 
  count & 871 &  99 &  38 &  13 &  26 \\ 
\bottomrule
\end{tabular}
\end{small}
\end{center}
\end{table}

\subsection{Case Linkage Results}
For the $946$ solved crimes in the testing period, we compared all $\binom{946}{2}=446{,}985$ crime pairs (564 linked and $446{,}421$ unlinked) by first transforming the crime variables into evidence variables and calculating the estimated Bayes factors from all models. 
For each model, the crime pairs were ordered according to their estimated Bayes Factor and the predictive performance is determined from the number of actual linkages found in the ordered list. 

\begin{figure}
\centering
\includegraphics[width=\textwidth]{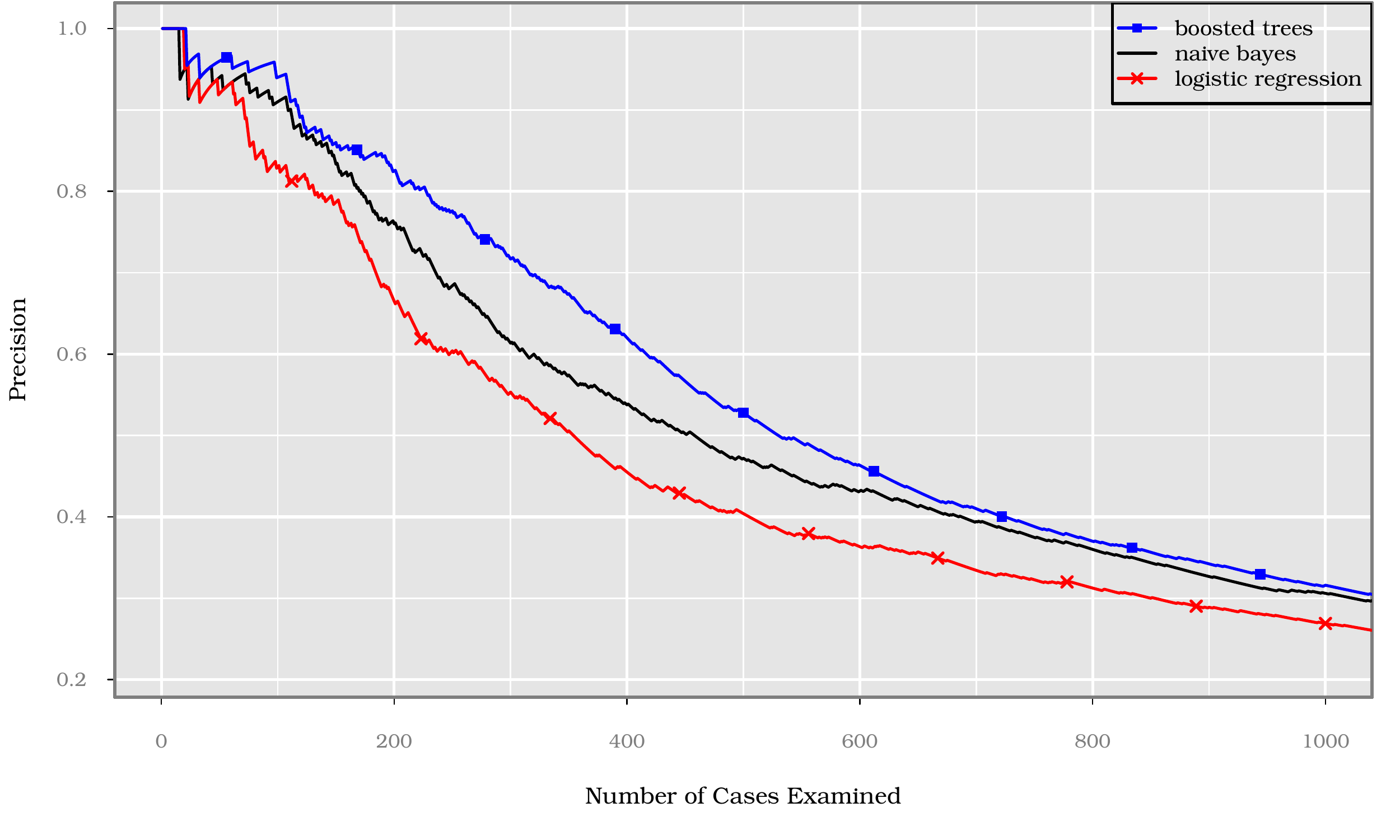} \\[10pt]
\includegraphics[width=\textwidth]{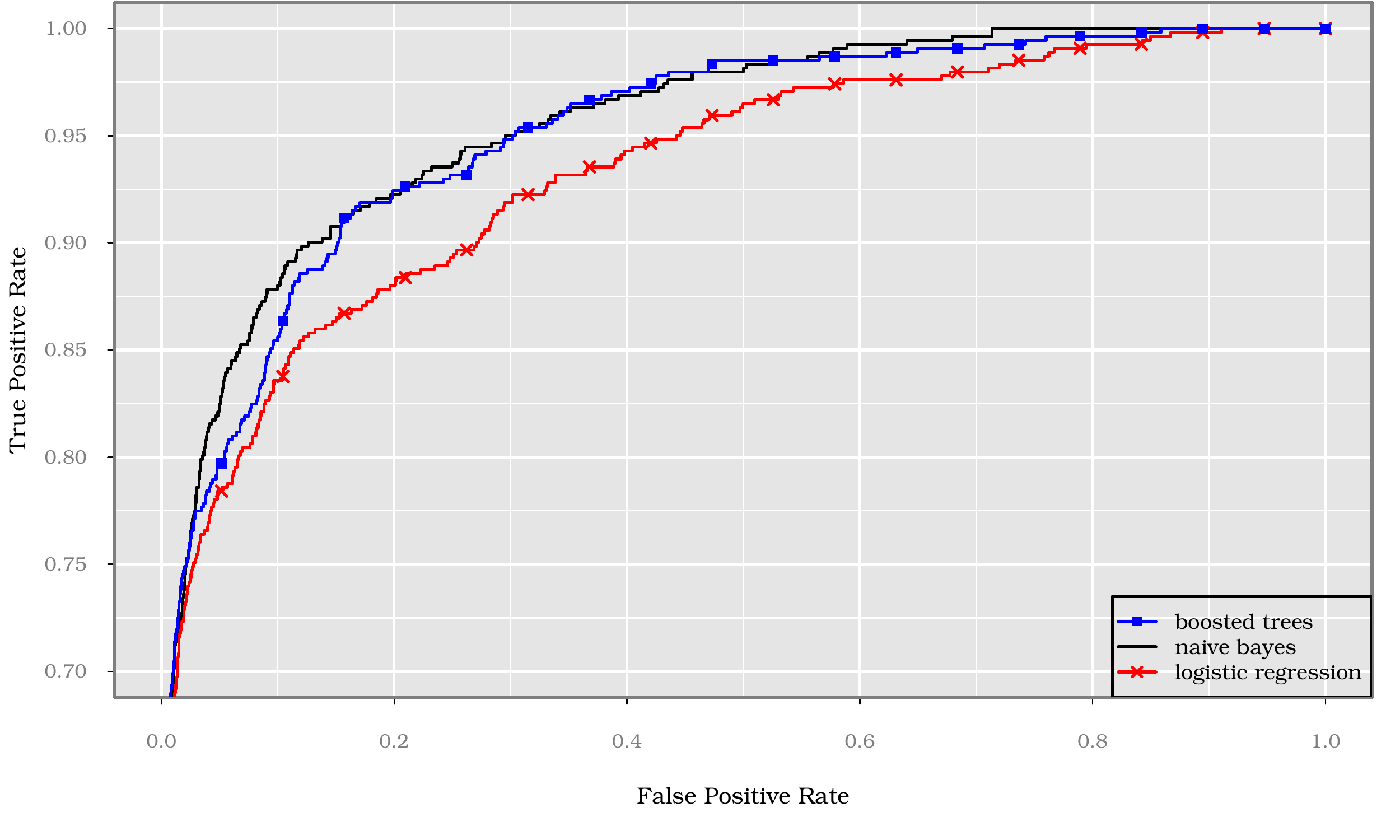}
\caption{Predictive performance of pairwise case linkage models. Top: Precision vs. workload curves. Bottom: ROC curves.}
\label{fig:ROC-caseLinkage}
\end{figure}

Two performance metrics were considered in Figure~\ref{fig:ROC-caseLinkage}. The top plot shows the proportion of the ordered crime pairs that are actual linkages (\emph{Precision}) for a given number of cases examined. 
Under the boosted trees model, for example, about 94/100 (94\%) of the highest ranked crime pairs are actually linked. 
ROC curves are shown in the bottom plot. This shows the proportion of all linked pairs (\emph{True Positive Rate}) that would be captured if a given proportion of unlinked pairs (\emph{False Positive Rate}) were examined. 
Under the na{\"i}ve Bayes model, for example, about 82\% of the actual linkages 
could be identified if a crime analyst was willing to sift through 5\% 
(or 22{,}321)
of the non-linked crime pairs. 

These results show that all models have comparable performance if only the crime pairs (e.g., $<50$) with the strongest evidence of linkage are evaluated. 
If more crime pairs are evaluated, the boosted trees and na{\"i}ve Bayes models 
perform better than logistic regression.
Because the na{\"i}ve Bayes model fits well, is simple to interpret, and quick to extract values, we use it going forward.
 
\subsection{Series Identification Results}
While the results from pairwise case linkage suggests that the Bayes factor can be a useful metric for determining linked crimes, the pairwise only comparisons may not be very helpful to crime analysts who are more interested in capturing all the crimes from a crime series.
Therefore, this section evaluates the performance of our approach to crime series identification.
There were 176 solved crime series with more than one event during the testing period. 
For each of these series, one event was removed and the remaining series was compared to the removed event as well as every other event (solved and unsolved) in the testing period. 
The 5048 unsolved crimes were included to better reflect a realistic implementation \citep{Woodhams-Labuschagne-2012,Tonkin-etal-2012-FSI}.
The log Bayes factor (from the na{\"i}ve Bayes model) was used as the similarity score for crime pairs and we considered single, complete, and average linkage for comparing each crime to a crime series.
This is intended to assess how well an additional crime from a crime series can be identified. 

Evaluation is based on the rank of score given to the removed crime.
If the removed crime was most similarity to the remaining series, then the rank was 1.
Table~\ref{t:seriesID} shows the proportion of series with rank in the top $R$.
Because unsolved crimes were also compared, it is possible that some unsolved crimes were actually (undetected) members of the crime series. Therefore, we also show the ranking when only solved crimes were included in the comparison.
For every crime series we held out each event in turn and returned the fraction of the held out events that are within the top $R$ results, ensuring long and short series are treated equally.

\begin{table}[!ht]
\caption{Summary of series identification performance using both solved and unsolved crimes (solved crimes only are given in parentheses). ``P rank R'' is the proportion of crime series whose crimes are associated with the true series with a similarity score in the top $R$ rank. The series size is calculated after removing one event for testing. The ranks are based on 5994 crimes, 946 of which are solved. }\label{t:seriesID}
\vspace{0.2in}
\begin{small}
\hspace{0.1in}(a) All series ($n=176$)
\begin{center}\begin{tabular}{l|ccccc}
 Linkage  & P rank 1 & P rank 5 & P rank 10 & P rank 25 & P rank 50 \\ 
  \hline
Single & 0.37 (0.50) & 0.54 (0.71) & 0.61 (0.78) & 0.74 (0.82) & 0.77 (0.89)  \\ 
Complete & 0.30 (0.45) & 0.50 (0.68) & 0.57 (0.76) & 0.69 (0.80) & 0.74 (0.87)  \\ 
Average & 0.37 (0.51) & 0.55 (0.71) & 0.62 (0.78) & 0.74 (0.82) & 0.77 (0.89) \\
\end{tabular}\end{center}

\hspace{0.1in}(b) One crime in series ($n=99$)
\begin{center}\begin{tabular}{l|ccccc}
 Linkage & P rank 1 & P rank 5 & P rank 10 & P rank 25 & P rank 50 \\ 
  \hline
Single & 0.26 (0.41) & 0.43 (0.62) & 0.49 (0.71) & 0.65 (0.76) & 0.69 (0.85)  \\ 
Complete & 0.26 (0.41) & 0.43 (0.62) & 0.49 (0.71) & 0.65 (0.76) & 0.69 (0.85)  \\ 
Average & 0.26 (0.41) & 0.43 (0.62) & 0.49 (0.71) & 0.65 (0.76) & 0.69 (0.85) \\ 
\end{tabular}\end{center}

\hspace{0.1in}(c) Two-three crimes in series ($n=51$)
\begin{center}\begin{tabular}{l|ccccc}
 Linkage & P rank 1 & P rank 5 & P rank 10 & P rank 25 & P rank 50 \\ 
  \hline
Single & 0.55 (0.66) & 0.70 (0.84) & 0.80 (0.88) & 0.87 (0.91) & 0.88 (0.92)  \\ 
Complete & 0.42 (0.55) & 0.65 (0.79) & 0.73 (0.85) & 0.80 (0.89) & 0.85 (0.90) \\ 
Average & 0.53 (0.66) & 0.71 (0.84) & 0.82 (0.89) & 0.87 (0.90) & 0.89 (0.94) \\ 
\end{tabular}\end{center}

\hspace{0.1in}(d) Four or more crimes in series ($n=26$)
\begin{center}\begin{tabular}{l|ccccc}
 Linkage  & P rank 1 & P rank 5 & P rank 10 & P rank 25 & P rank 50 \\ 
  \hline
Single & 0.45 (0.57) & 0.62 (0.80) & 0.70 (0.85) & 0.80 (0.91) & 0.85 (0.96)  \\ 
Complete & 0.22 (0.41) & 0.49 (0.70) & 0.58 (0.75) & 0.65 (0.80) & 0.72 (0.88) \\ 
Average & 0.48 (0.59) & 0.67 (0.82) & 0.74 (0.86) & 0.81 (0.92) & 0.84 (0.95) \\  
\end{tabular}\end{center}
\end{small}
\end{table}

While the overall probability of the correct cluster (P rank 1) is fairly low, the full model includes the correct cluster in the top five (P rank 5) 55\% of the time and the top fifty (P rank 50) 77\% of the time.  
Even in the challenging case when only a single crime is available, the additional crime is correctly attributed in 26\% of the cases and in the top fifty 69\% of the cases. 
Average linkage appears to be the best cluster similarity measure for series identification (in this example). 
Performance is best for crime series consisting of 2-3 crimes, but then slightly decreases for series longer than 4. 
The explanation for this decrease in performance is unclear. It may be a function of statistical fluctuation (there are only $n=26$ series with more than four crimes) or it may be that longer crime series are committed by offenders who are more difficult to link and consequently apprehend.

We also note that in addition to a ranked list of crimes that could be part of the series, this approach also provides the score (in the form of a Bayes factor) that is directly related to how confident we can be that the crime is linked to the series. This allows the analyst to include crimes to further investigate only when the score is sufficiently high.

\subsection{Crime Series Clustering Results}
\begin{figure}[!htbp]
\centering
\includegraphics[width=\textwidth]{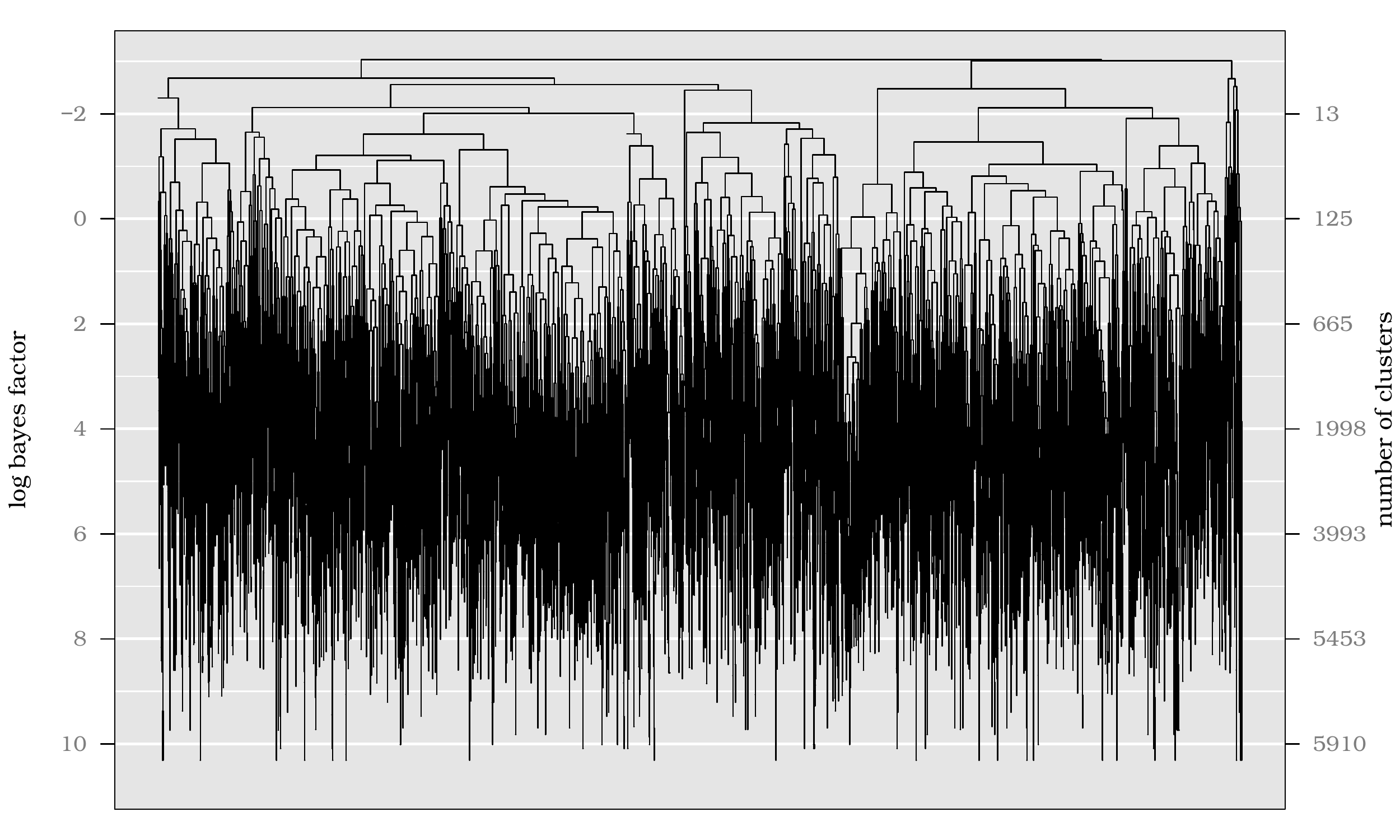}
\caption{Dendrogram from the average linkage agglomerative hierarchical clustering of all crimes (solved and unsolved) from the testing period.}
\label{fig:dendro}
\end{figure}

All 5994 crimes (both solved and unsolved) from the testing period were clustered using agglomerative hierarchical clustering with single, complete, and average linkage.
Figure~\ref{fig:dendro} shows the resulting dendrogram from average linkage hierarchical clustering.
This required making $\binom{5994}{2}=17{,}961{,}021$ pairwise comparisons using the Bayes Factor estimates from the na{\"i}ve Bayes model. 
We labeled any pair that involved an unsolved crime ``unknown" as it could not be determined if any of these pairs were linked. This resulted in 564 linked pairs, $446{,}421$ unlinked pairs, and $17{,}514{,}036$ unknown pairs. 

Table~\ref{t:treeResults} provides some evaluation metrics for the clusterings at various thresholds. The table shows the number of clusters if the dendrogram is cut at the corresponding log Bayes Factor
as well as the number of series (or number of clusters that consist of two or more crimes) that are present.
For performance, the number of linked pairs that are correctly contained in a series (out of 564 possible pairs), the number of unlinked pairs that are incorrectly assigned to the same cluster (out of 446{,}421), and the number of pairs with unknown linkage contained in a series are shown.
The variation of information (VI) metric \citep{Meila-2007} is an information-theoretic metric (based on the estimated conditional entropy) to evaluate the quality of the cluster solutions at a given threshold. Lower values indicate better performance and a perfect clustering solution will have a score of zero.

The average linkage approach results in the best performance as it includes the most linked pairs in the same cluster while limiting the number of unlinked pairs that are incorrectly assigned to the same cluster. 
As an example, using average linkage with a threshold of 7 (selected because it has the lowest VI) results in 821 estimated crime series in the testing period. These crime series contain 139 (24.6\%) of the linked pairs, but only mistakenly include 35 (0.008\%) of the unlinked pairs. 
At this particular threshold, the average linkage method assigns the remaining 4061 crimes into singleton clusters indicating that there is not enough evidence to link them with any other crimes.

\begin{table}[!htbp]
\begin{center}
\begin{small}
\begin{tabular}{cccccccc}
\toprule
& &  & & \# of & \# of & \# of & \\
& &  & & linked & unlinked & unknown & \\
& & \# of  & \# of    & pairs in & pairs in &  pairs in & \\
& log BF & clusters & series  &a series &  a series & a series & VI
\\
\midrule
\multirow{11}{*}{\rotatebox{90}{\textbf{Single Linkage}}} 
   & 6.0 & 2725 & 551 & 326 & 14701 & 433887 & 1.73 \\ 
   & 6.4 & 3605 & 698 & 278 & 1650 & 50097 & 0.76 \\ 
   & 6.8 & 4293 & 723 & 233 & 226 & 6438 & 0.45 \\ 
   & 7.2 & 4779 & 642 & 167 & 89 & 3121 & 0.40 \\ 
   & 7.6 & 5103 & 537 & 141 & 41 & 1696 & 0.39 \\ 
   & 8.0 & 5393 & 403 & 120 & 17 & 893 & 0.39 \\ 
   & 8.4 & 5570 & 292 & 90 & 5 & 600 & 0.41 \\ 
   & 8.8 & 5711 & 188 & 60 & 4 & 428 & 0.44 \\ 
   & 9.2 & 5850 & 102 & 38 & 2 & 184 & 0.46 \\ 
   & 9.6 & 5883 & 78 & 33 & 2 & 137 & 0.46 \\ 
   & 10.0 & 5908 & 66 & 24 & 2 & 99 & 0.47 \\   
   
\midrule 
\multirow{11}{*}{\rotatebox{90}{\textbf{Complete Linkage}}}     
   & 0.0 & 758 & 752 & 289 & 763 & 26491 & 0.88 \\ 
   & 1.0 & 1090 & 1069 & 287 & 485 & 17534 & 0.68 \\ 
   & 2.0 & 1494 & 1436 & 275 & 326 & 12059 & 0.58 \\ 
   & 3.0 & 1981 & 1782 & 257 & 228 & 8247 & 0.50 \\ 
   & 4.0 & 2576 & 1953 & 199 & 146 & 5662 & 0.46 \\ 
   & 5.0 & 3310 & 1840 & 179 & 100 & 3792 & 0.42 \\ 
   & 6.0 & 4132 & 1417 & 150 & 59 & 2382 & 0.41 \\ 
   & 7.0 & 4944 & 840 & 111 & 30 & 1245 & 0.40 \\ 
   & 8.0 & 5477 & 426 & 69 & 9 & 611 & 0.43 \\ 
   & 9.0 & 5820 & 135 & 37 & 2 & 204 & 0.46 \\ 
   & 10.0 & 5910 & 66 & 20 & 2 & 98 & 0.48 \\

\midrule 
\multirow{11}{*}{\rotatebox{90}{\textbf{Average Linkage}}}      
   & 0.0 & 125 & 119 & 373 & 9888 & 313370 & 2.57 \\ 
   & 1.0 & 320 & 307 & 350 & 3889 & 119359 & 1.76 \\ 
   & 2.0 & 665 & 618 & 321 & 1619 & 50237 & 1.17 \\ 
   & 3.0 & 1240 & 1076 & 304 & 725 & 21783 & 0.77 \\ 
   & 4.0 & 1998 & 1450 & 286 & 379 & 11253 & 0.57 \\ 
   & 5.0 & 2950 & 1578 & 269 & 180 & 5855 & 0.44 \\ 
   & 6.0 & 3993 & 1331 & 185 & 78 & 3023 & 0.40 \\ 
   & 7.0 & 4882 & 821 & 139 & 35 & 1433 & 0.39 \\ 
   & 8.0 & 5453 & 419 & 92 & 9 & 672 & 0.41 \\ 
   & 9.0 & 5816 & 133 & 41 & 2 & 212 & 0.45 \\ 
   & 10.0 & 5910 & 66 & 20 & 2 & 98 & 0.48 \\   
    
\toprule 
&  & &  \textbf{Totals:} & \textbf{564 }&\textbf{446{,}421} & \textbf{17{,}514{,}036} &  \\ 
\end{tabular}
\end{small}
\end{center}
\caption{Results from crime series linkage under the na{\"i}ve Bayes model. If the dendrogram is cut at a given log Bayes Factor, the table shows the number of clusters and the number of crime series (i.e., clusters with more than one crime) that would be produced. 
VI is the variation of information metric \citep{Meila-2007}.
 }
\label{t:treeResults}
\end{table}

\subsection{Suspect Prioritization Results}
This section tests the ability of our system to classify the offender of a crime or crime series. 
We considered the 4147 known offenders active during the training period as the possible suspects and recorded their crime series from the crimes that were committed during the training period only.
Next, we compared each of the 1047 crime series from the testing period to all of the series from the training period.
There were 109 offenders that were arrested for crimes in both the training and testing periods. 

Because some of the suspect's past crime series occurred years before the testing period we don't include the temporal proximity (\texttt{temporal}) variable in the model. 
This is easy to accommodate in the na{\"i}ve Bayes model by simply removing the temporal predictor from the model as no re-calibration is necessary. 

\begin{figure}[!ht] 
\centering
\includegraphics[width=\textwidth]{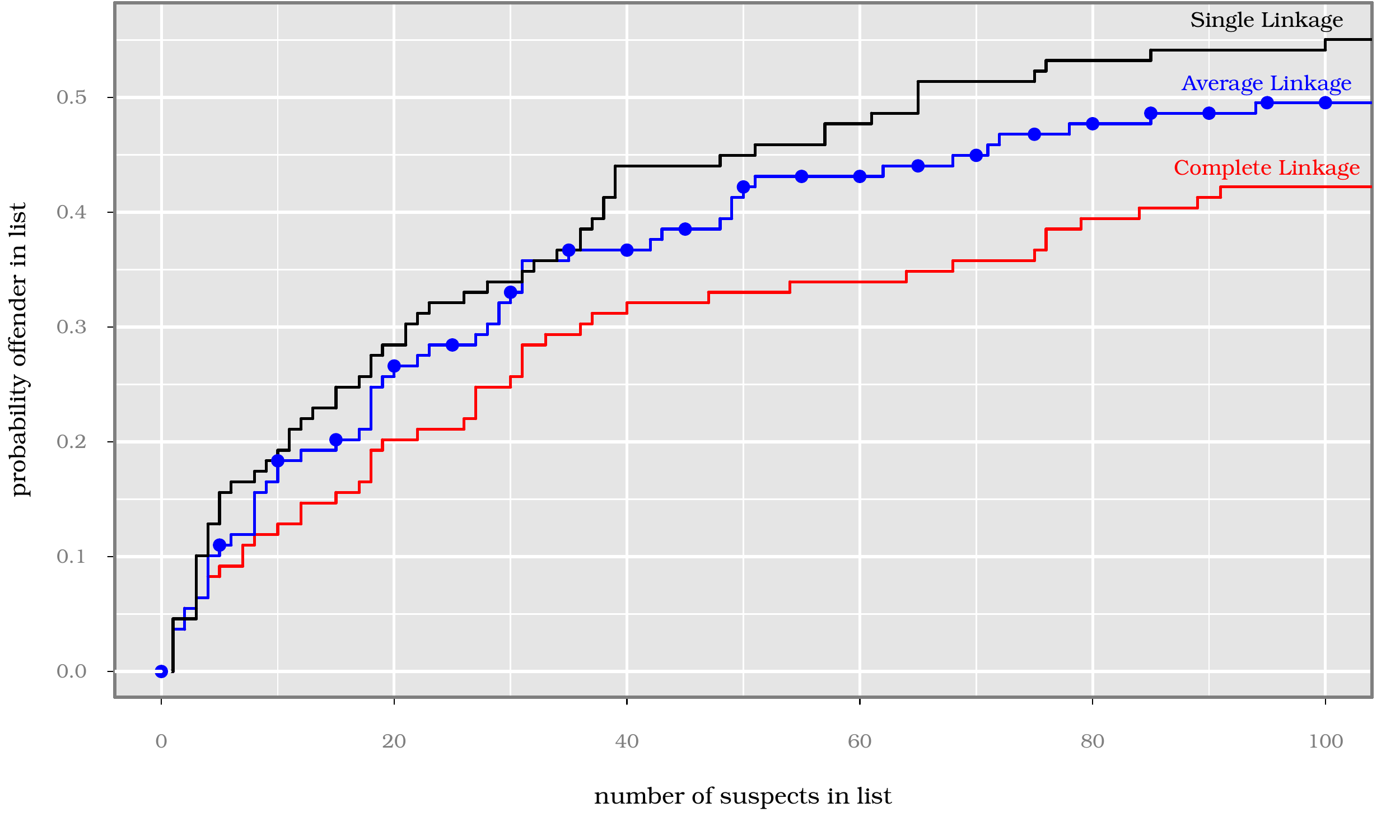}
\caption{Performance of suspect prioritization for single, complete, and average linkage for offenders that have crimes in both the training and testing periods. 
This plots the proportion of offenders that would be contained in a ranked list of the 4147 suspects.} 
\label{fig:susPri}
\end{figure}

For a given crime series from the testing set, we compared the crimes in the series to all the crime series from the training period using single, complete, and average linkage (using a na{\"i}ve Bayes model with the temporal proximity component removed). 
Suspects are ranked according to the similarity score between their crime series during the training period and the crimes from the current series under consideration.
For the 109 crime series where the offender is in the suspect list, 
Figure~\ref{fig:susPri} shows the proportion of the offenders that would be contained in a ranked list of suspects under single, complete, and average linkage. For this scenario, the single linkage technique performs best.

However, there is no guarantee the actual offender is in the list of suspects.
Because of time and manpower constraints, police investigators may only be able to investigate (e.g., find and interview) suspects when there is strong evidence that they are responsible.
Therefore, it is important to provide not only a ranked list of suspects but also the score (i.e., the estimated log Bayes Factor) between the current series and the suspects' past crime series so investigators can invest their resources only when there is sufficient evidence to do so. 

Table~\ref{t:suspectPrioritization} provides more details on the performance of this method of suspect prioritization by showing the estimated probability that the true offender will be included in the short list of suspects when investigations are made only when the estimated log Bayes Factor exceeds a certain threshold.
As an example, for a threshold of 3 around 30 suspects will be candidates for investigation. If one of the suspects is responsible for the new crime series, they will be on the list in about 40\% of the cases. 
However, because only 10.4\% (109/1047) of the offenders in the testing period are suspects (i.e., were arrested for crimes in the training period), the overall probability that an offender can be identified is closer to only $0.04$.

While substantially better than random guessing, the performance of these models for suspect prioritization is not exceptional.
It is furthermore noted that the reported performance is likely inflated as the crimes series from the testing period were not estimated but extracted from the solved crimes and thus known without error (according to police records).
While it is possible to increase the probability that an offender is in the suspect list by including past offenders that were active prior to the training period or that committed other types of crimes, it is not clear if this will lead to improved performance as it will also provide a higher false alarm rate.

\begin{table}[ht]
\centering
\begin{tabular}{crrrrrrr}
\toprule 
&		   & \multicolumn{6}{c}{log Bayes Factor} \\ 
& Threshold &  \multicolumn{1}{c}{0} & \multicolumn{1}{c}{1} & \multicolumn{1}{c}{2} & \multicolumn{1}{c}{3} & \multicolumn{1}{c}{4} & \multicolumn{1}{c}{5} \\ 
\midrule 
\multirow{4}{*}{\parbox{2.5cm}{\centering Number of suspects in list}} & 
  1st Quar &  352.00 & 119.00 & 42.00 & 14.00 & 4.00 & 0.00 \\ 
&  Median & 500.00 & 194.00 & 76.00 & 30.00 & 9.00 & 2.00 \\ 
&  Mean  & 544.30 & 221.50 & 90.82 & 37.34 & 12.99 & 4.40 \\ 
&  3rd Quar & 725.00 & 310.50 & 128.50 & 51.00 & 17.00 & 6.00 \\
\cline{1-8} 
\multirow{2}{*}{\parbox{2.5cm}{\centering Probability offender in list}}  
&  Conditional  & 0.75 & 0.62 & 0.56 & 0.40 & 0.25 & 0.13 \\ 
&  Overall & 0.08 & 0.06 & 0.06 & 0.04 & 0.03 & 0.01 \\ 
\bottomrule
\end{tabular}
\caption{Performance of suspect prioritization for single linkage when suspects are investigated only when the score (i.e., log Bayes factor) between their past crimes and the current crime series exceeds a certain threshold. 
The probability the true 
offender's series exceeds the threshold 
is given conditional on the offender actually being included in the suspect list as well as the overall probability that assumes there is a 10.4\% chance the true offender is in the suspect list.  
}
\label{t:suspectPrioritization}
\end{table}

\section{Discussion} \label{sec:discussion}
It was our aim to assist crime analysts by developing statistical methodology that provides more accurate crime linkage assessments and predictions.
The interpretable statistical models offer the potential to put linkage analysis on a more reliable and scientific basis, increase standardization, reduce the cognitive workload placed on the analyst, and improve the prospect of using linkage analysis as evidence in legal proceedings. 
 We derived and proposed the use of the Bayes Factor for making pairwise case linkage decisions. Grounded in Bayesian decision theory, the Bayes factor is the ratio between the similarity and distinctiveness of a pair of crimes and a direct measure of the strength of evidence favoring the linkage hypothesis. 
	
Logistic regression, na{\"i}ve Bayes, and boosted trees were used to carry out pairwise case linkage on 446{,}985 breaking and entering crime pairs using the spatial and temporal proximity between crimes as well as several MO characteristics.
We did not focus on finding the optimal tuning parameters, transformation functions, or classification models.
We expect performance would be increased through the use of more sophisticated models tuned to specific policing jurisdictions.

Using the pairwise Bayes factors from case linkage and methods from hierarchical clustering, we presented an approach to find the additional events from a crime series (crime series identification). 
This represents the scenario for a crime analyst tasked with examining a new crime (e.g., reported during the previous night) and attempting to assign it to an existing series.
Our method produces a ranked list of candidate crimes according to how likely they are part of the existing crime series.
The average linkage method could identify about 77\%-89\% of the true additional crimes from the series in a list of the top 50 crimes.
If there were two or three crime already in the series, the performance improved to 89\%-94\%.

We further examined how well our method can identify the offender of a crime series (suspect prioritization). 
If suspects are only investigated when the log Bayes Factor between their past crime series and the current series exceeds 3.0 (under single linkage), then around 30 suspects must be investigated to achieve a 40\% chance of finding the true offender, when they are in the suspect list.
But because there is a low probability that the offender was actually in the suspect list, the unconditional probability that this method can find the true offender is only around 4\%.
We speculate that including the suspects' home location and incorporating methods from geographic profiling has the potential to significantly improve suspect prioritization. 

In practice, crime analysts use a variety of methods to make linkage decisions. 
Even if the performance of our methods alone are not sufficient for a particular scenario, they may still be informative to the decision making process by providing additional insight into the data or used in conjunction with other methods. 
However, we have demonstrated the ability of our models to make correct linkages using a variety of real-world scenarios and actual crime data. 
Importantly, our models also provide a method to evaluate the strength of the linkage assessment,
allowing a crime analysis to act only when sufficient evidence exists. 
Furthermore, the statistical framework provides a foundation to examine the assumptions that linkage analysis rests on, facilitating a standard approach for comparing the results for different crime types across different locations. 

The quality of the police data will limit the performance of any linkage methodology. 
However, by using raw police data our findings reflect realistic performance under the same conditions that an analyst would experience. 
The sensitivity of our models to data quality issues is difficult to assess without the analysis of more data from different sources. The ability to make correct linkage decisions depends on both the consistency of the offenders' behaviors and the distinctiveness of their crimes relative to the other offenders in the jurisdiction \emph{as recorded in the crime data}. Poor data quality will make it appear to the linkage system that the offenders' behaviors are either not as consistent or distinctive as they are in reality. As a result, linkage performance will suffer.
\citet{Bouhana-etal-2014} offers an approach for measuring how well the crime data can discriminate between serial offenders.

All models were implemented in R \citep{R}. Code is available from the  author upon request.

\clearpage
\singlespacing
\bibliographystyle{asa}
\newpage
\bibliography{linkage}

\end{document}